\documentclass[aps,prb,twocolumn,groupedaddress]{revtex4}  
\usepackage{graphicx}  
\usepackage{dcolumn}   
\usepackage{bm}        
\usepackage{amssymb}   
\usepackage{amsmath}
\usepackage{dsfont}
\usepackage{mathrsfs}

\begin{document}

\newcommand{\bra}[1]{\left\langle#1\right|}
\newcommand{\ket}[1]{\left|#1\right\rangle}
\newcommand{\ev}[1]{\langle#1\rangle}
\newcommand{\cstate}[4]{\ket{\begin{smallmatrix}#1 & #2\\#3 & #4\end{smallmatrix}}}
\newcommand{\Tr}[1]{\mathrm{Tr}\left(#1\right)}

\title{Entanglement Dynamics of Molecular Exciton States in Coupled Quantum Dots}
\author{Cameron Jennings}
\author{Michael Scheibner}
\affiliation{School of Natural Sciences, University of California Merced, Merced, California 95343, USA}
\date{\today}

\begin{abstract}
We theoretically model the electronic dynamics of a coupled quantum dot pair in a static electric field. We then investigate the possibility of polarization-entangled photon emission from the radiative cascade of the molecular biexciton state. Through numerical simulations, we analyze the dependence of entanglement fidelity on temperature and electric field, as well as tunnel coupling. We establish a regime of direct-indirect exciton detunings for which coupled quantum dots are superior to single dots for entangled photon generation, yielding near-unit fidelity over a larger range of exchange splittings.
\end{abstract}

\maketitle

\section{Introduction}

Quantum mechanics allows for the unique possibility of instantaneous non-local correlation, or entanglement, between particles. Several technologies have emerged which use quantum entanglement as a fundamental resource, including quantum computation\cite{Feynman1982,Deutsch1985,Nielsen2000,Knill2001,Nielsen2004,Prevedel2007} and cryptography,\cite{Nielsen2000,Prevedel2007,Ekert1991,Bennett1992,Gisin2002,Walton2003} as well as entangled-photon microscopy.\cite{Ono2013} Quantum information technologies can be implemented by using the orthogonal polarization states of a photon to encode binary information in a ``dual-rail'' qubit scheme,\cite{vanLoock2011} which allows for simple manipulation and distribution of photonic qubits using linear optical elements,\cite{Kok2007} or using spin states of bound charges.\cite{Economou2012}  However, a source of entanglement between photons or distant spins is needed to implement two-qubit gates or two-photon quantum cryptography protocols.

Entangled photon pairs are routinely produced by nonlinear optical processes such as spontaneous parametric downconversion (SPDC), in which one photon interacts with a bulk crystal to produce two lower-energy photons with opposite polarizations. However, due to the probabilistic nature of such an interaction, SPDC-based entangled photon sources can always produce zero, one, or multiple photon pairs from each incident pulse, which can introduce errors in entanglement-based quantum information.\cite{Dusek2002} Semiconductor quantum dots (QDs) are promising candidates for on-demand sources of entangled photon pairs, as well as spin-photon entanglement.\cite{DeGreve2012,Gao2012} Two-photon emission can be achieved by optically exciting the neutral biexciton state, consisting of two bound excitons. Radiative decay from this state can proceed along one of two pathways corresponding to the two optically active single-exciton spin states. Coherence between these two oppositely-polarized (but otherwise indistinguishable) emission processes results in a pair of polarization-entangled photons. In single QDs, however, the degeneracy of the intermediate exciton spin states is often removed by the anisotropic electron-hole exchange splitting, which is due to asymmetry in the self-assembled QD system.\cite{Bayer2002,Xu2008,Tong2011} If this splitting is larger than the optical transition linewidths, the emitted photons are distinguishable by their energy and exhibit correlated---but not entangled---polarizations.\cite{Santori2002} Several techniques have been implemented to avoid this problem, including postselection via spectral filtering\cite{Akopian2006} and minimization of the exchange splitting by postgrowth thermal annealing\cite{Greilich2006} or application of external fields.\cite{Trotta2012,Plumhof2012} Alternative QD growth methods have also been demonstrated to exhibit a systematically sma\-ll\-er exchange splitting, including interface fluctuation QDs\cite{Ghali2011} and QDs grown on the (111) surface of GaAs.\cite{Juska2013,Kuroda2013} Aside from neutral biexciton decay, negatively-charged states in QDs can also be used to create a chain of entangled photons in a one-dimensional cluster state.\cite{Lindner2009} This procedure can be generalized to two dimensions using two vertically-stacked coupled quantum dots (CQDs).\cite{Economou2010} Here we discuss a scheme in which entangled photon pairs are obtained straightforwardly through a cascaded biexciton-exciton decay.

Our scheme for on-demand entangled photon pairs uses CQDs grown in an electric field-effect structure, such as a Schottky diode.\cite{Scheibner2012,Ramirez2010,Skold2013,Bester2005} The CQD structure permits two types of neutral excitons: direct excitons, with the electron and hole in the same dot, and indirect excitons, with the electron and hole separated in different dots. Because of their spatial charge separation, the energies of indirect states can easily be tuned by applying an electric field.\cite{Kerfoot2014,Ramanathan2013,Szafran2007} In addition, spatial charge separation greatly reduces both the short-range isotropic and long-range anisotropic exchange splitting of indirect exciton spin states.\cite{Skold2013,Scheibner2007} This property can potentially be used for generation of entangled photon pairs by preparing the CQD in an indirect biexciton spin singlet state, with one direct (single-dot) exciton and one indirect exciton. In an effort to determine the utility of this proposal, we theoretically calculate the fidelity of entangled photon generation in a CQD system.

\begin{figure*}
\centering
\includegraphics[width=\textwidth]{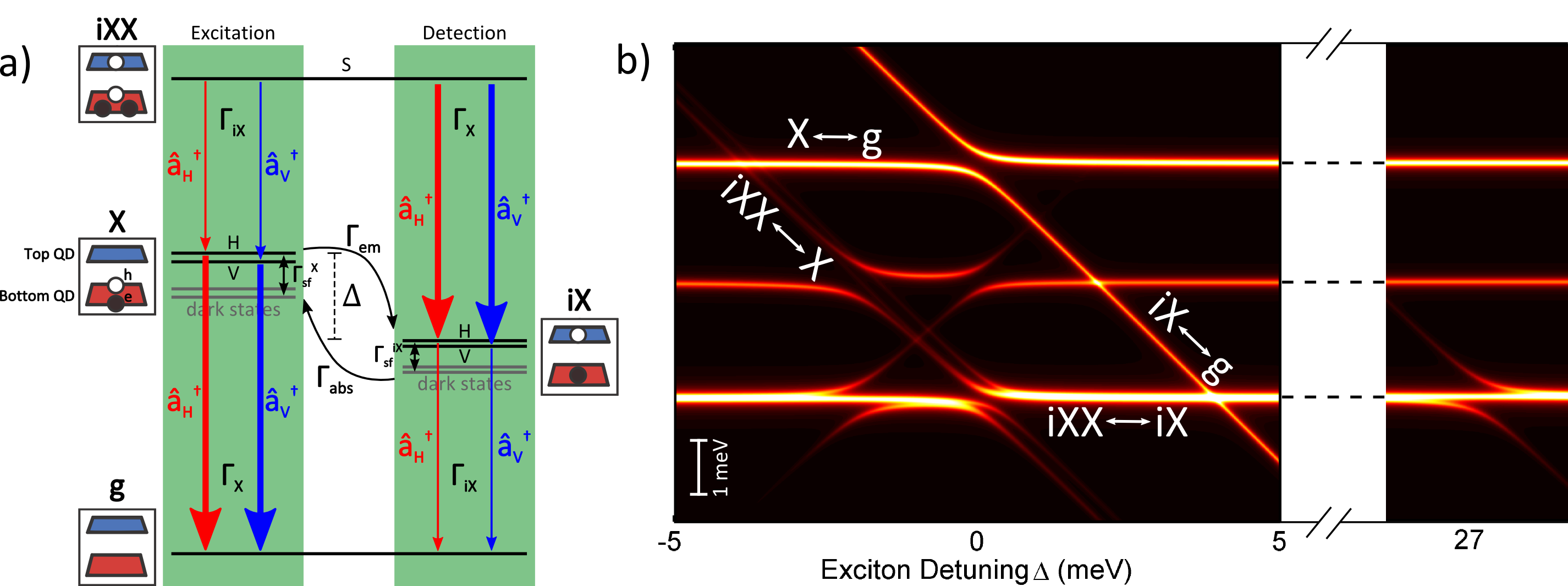}
\caption{(a) Energy level diagram of the molecular biexciton radiative cascade, depicting polarizations of all allowed recombination pathways, as well as spin-flip and phonon-assisted hole tunneling channels. Thick vertical arrows indicate direct exciton recombination, while thin vertical arrows indicate a slower interdot recombination. Excitation and detection pathways are indicated by green shading. (b) Simulated electric field-dispersed photoluminescence spectrum, mapping out each of the relevant transition energies in the vicinity of the exciton and biexciton hole tunneling resonances. Single-exciton hole tunneling resonance occurs at $\Delta=0~$meV, while biexciton hole tunneling resonances occur at $\Delta=-1.9$ and $27.2~$meV.}
\label{biexciton cascade}
\end{figure*}

\section{Theoretical Model}\label{model}

We consider an asymmetric CQD system of two vertically stacked, tunnel-coupled, self-assembled InAs-GaAs QDs (referred to here as the bottom and top dots) with respective heights $h_B$ and $h_T$ and center-to-center separation $d$, grown in a diode structure to allow application of an electric field $F$ along the growth direction. Depending on the dot asymmetry and applied field direction, either electron or hole levels can be tuned into resonance, causing electrons or holes to tunnel coherently.\cite{Bracker2006} For the remainder of this work and without loss of generality we will assume a CQD diode structure is chosen that promotes hole tunneling, with any excited electrons confined to the bottom dot.

We restrict the CQD Hilbert space to that of the crystal ground state $\ket{g}$, direct exciton $\ket{X}$, indirect exciton $\ket{iX}$, and molecular biexciton $\ket{iXX}$ (consisting of one direct and one indirect exciton). Each of the single-exciton charge states can exist in one of four spin states: two optically active bright states $\ket{H/V}=\ket{\downarrow\Uparrow \pm \uparrow\Downarrow}$, and two optically inactive dark states $\ket{H_d/V_d}=\ket{\uparrow\Uparrow \pm \downarrow\Downarrow}$ (single (double) arrows denote spin-1/2 electrons (spin-3/2 heavy holes)).
The molecular biexciton can exist in a singlet $\ket{S}=\ket{\uparrow\downarrow-\downarrow\uparrow}\ket{\Uparrow\Downarrow-\Downarrow\Uparrow}$ or one of three triplet spin states $\ket{T_+}=\ket{\uparrow\downarrow-\downarrow\uparrow}\ket{\Uparrow\Uparrow}$, $\ket{T_0}=\ket{\uparrow\downarrow-\downarrow\uparrow}\ket{\Uparrow\Downarrow+\Downarrow\Uparrow}$, or $\ket{T_-}=\ket{\uparrow\downarrow-\downarrow\uparrow}\ket{\Downarrow\Downarrow}$.

While a CQD exhibits many optical transitions, we focus on the molecular biexciton spin singlet state $\ket{iXX,S}$, which can produce correlated photon pairs by sequential recombination to the ground state due to the optical selection rules. As depicted in Fig.~\ref{biexciton cascade}(a), this state is optically coupled to the ground state through the bright spin states of the direct and indirect exciton, forming two pairs of oppositely-polarized decay pathways.

Resonant excitation can selectively prepare the system in the molecular biexciton state. In contrast to biexcitons in single QDs, the molecular biexciton can be resonantly excited by stepwise two-laser driving along one of two pathways, via the direct or indirect exciton state. Since the indirect exciton pathway features a much smaller exchange splitting, the preferred two-laser excitation scheme is to drive each transition of the direct exciton pathway and detect the emission along the indirect exciton pathway, as depicted in Fig.~\ref{biexciton cascade}(a). The relevant transitions are identified in the simulated electric field-dispersed photoluminescence spectrum of Fig.~\ref{biexciton cascade}(b). The singlet spin state can be spectrally selected near a biexciton tunneling resonance, with triplet states excluded from the spin-conserving tunnel coupling, or by utilizing two-photon absorption, due to the selection rules of simultaneous transitions.\cite{Scheibner2015}

We consider the case of pulsed excitation, which allows for on-demand preparation of the molecular biexciton state and subsequent emission of a single entangled photon pair. For two-laser driving, one must further specify the relative timing of the two pulses: they may be applied sequentially, or concurrently. Bensky et al.\cite{Bensky2013} examined each of these pulsed excitation schemes in the case of a single QD and found that both can generate the biexciton state with more than $99\%$ probability, though the concurrent pulse scheme may be less reliable due to interference between the pulses. Pulsed excitation of the molecular biexciton has not been examined experimentally, though continuous-wave excitation has been demonstrated.\cite{Scheibner2015}

Photon emission events can be described within the CQD and photon subspaces using transition operators $\hat{\sigma}_{1H/V,D/I}$ ($\hat{\sigma}_{2H/V,D/I}$) and photon creation operators $\hat{a}^{\dagger}_{H/V}$, which act on the system to produce the first (second) $H/V$-polarized photon by direct/indirect recombination of the biexciton (exciton) state (see Table I for details). Assuming that the system has been prepared in the molecular biexciton spin singlet state $\ket{\Psi_0}=\ket{iXX,S}$ and that recombination occurs via the indirect exciton pathway without nonradiative transitions, the first recombination produces the state
\begin{equation}
\begin{aligned}
\ket{\Psi_1} &= \frac{1}{\sqrt{2}} (\hat{a}^{\dagger}_H \hat{\sigma}_{1H,D} + \hat{a}^{\dagger}_V \hat{\sigma}_{1V,D}) \ket{\Psi_0} \\
&= \frac{1}{\sqrt{2}} (\ket{H}_1 \ket{iX,H} + \ket{V}_1 \ket{iX,V}).
\end{aligned}
\end{equation}
During the time $\tau$ before the second photon is emitted, the system undergoes coherent evolution, acquiring a phase difference due to the exchange interaction:
\begin{equation}\label{intermediatestate}
\begin{aligned}
\ket{\Psi_1 (\tau)} &= e^{-i\hat{H} \tau / \hbar} \ket{\Psi_1(0)} \\
&= \frac{1}{\sqrt{2}} (\ket{H}_1 \ket{iX,H} + e^{i S_I \tau / \hbar} \ket{V}_1 \ket{iX,V}),
\end{aligned}
\end{equation}
to within a global phase factor, where $S_I$ is the exchange splitting between indirect exciton bright spin states. Finally, the second recombination results in
\begin{equation}
\begin{aligned}
\ket{\Psi_2} &= \frac{1}{\sqrt{2}} (\hat{a}^{\dagger}_H \hat{\sigma}_{2H,I} + \hat{a}^{\dagger}_V \hat{\sigma}_{2V,I}) \ket{\Psi_1 (\tau)} \\
&= \frac{1}{\sqrt{2}} (\ket{H}_1 \ket{H}_2 + e^{i S_I \tau / \hbar} \ket{V}_1 \ket{V}_2) \ket{g} \\
&\equiv \ket{\Psi_{2P}} \ket{g},
\end{aligned}
\end{equation}
with the two photons in a pure polarization-entangled state $\ket{\Psi_{2P}}$. However, since recombination can occur via multiple pathways, the result is a mixed state, described by the two-photon density matrix $\hat{\rho}_{2P}$.

In the absence of anisotropic electron-hole exchange splitting, each pair of oppositely-polarized decay pathways are energetically indistinguishable, resulting in the maximally-entangled Bell state $\ket{\Psi^+}=(\ket{H}_1 \ket{H}_2 + \ket{V}_1 \ket{V}_2)/\sqrt{2}$ for the polarizations of the emitted photon pair.\cite{Hafenbrak2007,Hudson2007} Under generic conditions, we quantify the quality of entanglement by the fidelity of the two-photon polarization state to the target Bell state. This fidelity is determined by the projection
\begin{equation}
\begin{aligned}
F^+&=\bra{\Psi^+}\hat{\rho}_{2P}\ket{\Psi^+} \\
&=\frac{1}{2}(\rho_{\ket{HH}\bra{HH}}+\rho_{\ket{VV}\bra{VV}})+\mathrm{Re}(\rho_{\ket{HH}\bra{VV}}).
\label{fidelity}
\end{aligned}
\end{equation}
A fidelity higher than $0.5$ indicates polarization entanglement, while a lower fidelity is acheivable by classical correlation alone. Uncorrelated photons, such as background light not originating from the CQD, give a baseline fidelity of $0.25$.\cite{Hudson2007,Bennett2010}

The CQD system interacts with the finite-temperature phonon bath of the surrounding crystal lattice, as well as the vacuum electromagnetic field via spontaneous emission. We assume Markovian behavior, such that the environment always remains at equilibrium (emitted particles cannot be reabsorbed). This assumption requires that any system-environment correlations decay on a much faster time scale than the system dynamics, which could introduce some inaccuracies at the lowest temperatures simulated ($T=1~$K) due to long-lived reservoir correlations. Using the Rotating Wave Approximation to average over quickly-oscillating terms, we can then describe the dissipative dynamics of this open quantum system using the Lindblad master equation\cite{Carmichael2002} for the reduced density matrix $\hat{\rho}$ of the system:
\begin{equation}\label{masterequation}
\frac{\partial}{\partial t}\hat{\rho}=\frac{1}{i\hbar}[\hat{H},\hat{\rho}] + \sum_i \Gamma_i \left( \hat{\sigma}^{~}_i\hat{\rho}\hat{\sigma}^{\dagger}_i - \frac{1}{2}\hat{\sigma}^{\dagger}_i\hat{\sigma}^{~}_i\hat{\rho} - \frac{1}{2}\hat{\rho}\hat{\sigma}^{\dagger}_i\hat{\sigma}^{~}_i \right)
\end{equation}

\begin{table}
\caption{List of incoherent processes included in the master equation, with corresponding transition rates and operators. Exciton spin states are indexed by $m=H,V,H_d,V_d$.}
\setlength{\extrarowheight}{3pt}
\begin{tabular}{c c c}
\hline
\hline
Process & Rate & Operator \\
\hline
Recombination & $\Gamma_X$ & $\hat{\sigma}_{1H/V,D}=\ket{iX,H/V}\bra{iXX,S}$ \\
 & & $\hat{\sigma}_{2H/V,D}=\ket{g}\bra{iX,H/V}$ \\
 & $\Gamma_{iX}$ & $\hat{\sigma}_{1H/V,I}=\ket{X,H/V}\bra{iXX,S}$ \\
 & & $\hat{\sigma}_{2H/V,I}=\ket{g}\bra{iX,H/V}$ \\
Phonon-assisted tunneling & $\Gamma_{abs}$ & $\ket{X,m}\bra{iX,m}$ \\
 & $\Gamma_{em}$ & $\ket{iX,m}\bra{X,m}$ \\
Phonon-assisted spin-flip & $\Gamma^X_{sf}$ & $\ket{X,H_d/V_d}\bra{X,H/V}$, H.c. \\
 & $\Gamma^{iX}_{sf}$ & $\ket{iX,H_d/V_d}\bra{iX,H/V}$, H.c. \\
Pure dephasing & $\gamma^X_p$ & $\sum_m \ket{X,m}\bra{X,m}$ \\
 & $\gamma^{iX}_p$ & $\sum_m \ket{iX,m}\bra{iX,m}$ \\
\hline
\hline
\end{tabular}
\label{incoherent}
\end{table}

We assume that an initial state of $\hat{\rho}(0)=\ket{iXX,S}\bra{iXX,S}$ is prepared by pulsed excitation, and do not explicitly include interaction with the optical pulse. By projecting onto the basis states, Eq.~\eqref{masterequation} can be written as $\bra{i}\dot{\hat{\rho}}\ket{j} \equiv \dot{\rho}_{ij}=\sum_{kl} M_{ij,kl} \rho_{kl}$, where the elements of the time-dependence tensor $M_{ij,kl}$ contain the various coherent phase evolution, transition and decay rates. Vectorizing the elements of $\hat{\rho}$ results in the matrix differential equation $\dot{\vec{\rho}}=M\vec{\rho}$, leading to the solution $\vec{\rho}(t)=e^{Mt}\vec{\rho}(0)$. In terms of the superoperator $\mathscr{L}\hat{\rho}=\partial_t \hat{\rho}$ defined by Eq.~\eqref{masterequation}, this solution can also be written as $\hat{\rho}(t) = e^{\mathscr{L}\tau} \hat{\rho}(0)$. Details of the numerical simulation are given in the appendix, including calculation of single-particle wavefunctions, Hamiltonian matrix elements, and transition rates.

For a given decay pathway, with the first (second) photon generated by $\alpha (\beta) = D/I$ recombination after a time delay $\tau$, the density matrix element $\bra{ij}\hat{\rho}_{2P}(\tau)\ket{kl}$ in the two-photon linear polarization basis is related to the CQD dynamics by the two-time correlation function\cite{Troiani2006,Carmele2010}
\begin{equation}
\begin{aligned}
g^{(2)}_{ijkl,\alpha\beta}(\tau) &= \ev{\hat{\sigma}^{\dagger}_{1i,\alpha}(0) \hat{\sigma}^{\dagger}_{2j,\beta}(\tau) \hat{\sigma}_{2l,\beta}(\tau) \hat{\sigma}_{1k,\alpha}(0)} \\
&= \mathrm{Tr}\left[ \hat{\sigma}_{2l,\beta} \, e^{\mathscr{L}\tau}( \hat{\sigma}_{1k,\alpha} \hat{\rho}(0) \hat{\sigma}^{\dagger}_{1i,\alpha} ) \, \hat{\sigma}^{\dagger}_{2j,\beta} \right] ,
\end{aligned}
\label{correlations}
\end{equation}
where the last line follows from the quantum regression theorem.\cite{Carmichael2002} These polarization correlations can be measured experimentally using time-correlated single-photon counting in a Hanbury Brown-Twiss setup,\cite{HanburyBrownTwiss1956} where each detection pathway contains polarization optics and a monochromator to spectrally select the relevant transitions and perform polarization cross-correlation measurements.\cite{Akopian2006,Hafenbrak2007} Since recombination can occur via several pathways with a random time delay, the elements of $\hat{\rho}_{2P}$ are calculated by using a time-averaged statistical mixture
\begin{equation}
\rho_{\ket{ij}\bra{kl}}=A\sum_{\alpha,\beta=D,I} \int_0^{T_d} d\tau\, P_{1\alpha}P_{2\beta}(\tau) g^{(2)}_{ijkl,\alpha\beta}(\tau),
\label{tomography}
\end{equation}
where $T_d$ is the detection time window, $A$ is a normalization constant set to enforce the condition $\Tr{\hat{\rho}_{2P}}=1$, $P_{1D(I)}=\Gamma_{X(iX)}/(\Gamma_{X}+\Gamma_{iX})$ is the probability of direct (indirect) recombination of the first exciton, $P_{2D(I)}(\tau)=\Gamma_{X(iX)} n_{X(iX)}(\tau)$ is the probability per unit time of the second direct (indirect) recombination, and $n_{X(iX)}(\tau)$ is the population of bright states $\ket{X(iX),\pm 1}$ at time $\tau$. Transition rates for direct and indirect exciton recombination are denoted by $\Gamma_X$ and $\Gamma_{iX}$, respectively.

The critical dynamics determining photon entanglement occurs during the time $\tau$ between photon emission events, when the dynamics is limited to the single-exciton subspace. Truncating to the bright spin states, the simplified exciton Hamiltonian can be written
\begin{equation}
\hat{H} =
\bordermatrix{
~ & \ket{iX,H} & \ket{iX,V} & \ket{X,H} & \ket{X,V} \cr
~ & S_I/2 & 0 & t_h & 0 \cr
~ & 0 & -S_I/2 & 0 & t_h \cr
~ & t_h & 0 & \Delta+S_D/2 & 0 \cr
~ & 0 & t_h & 0 & \Delta-S_D/2 \cr
},
\end{equation}
where $\Delta=ed(F-F_0)$ is the exciton detuning due to the applied electric field, $F_0$ is the field value at the $\ket{X}-\ket{iX}$ anticrossing, and $t_h$ is the resonant hole tunnel coupling including Coulomb correction. The magnitude of the exchange splitting $S_{D/I}$ is a critical factor for entangled photon emission, as a splitting larger than the radiative linewidth of the corresponding exciton is expected to render the two decay pathways distinguishable and prevent entanglement. The bright-state exchange splitting is approximated using the dipole interaction term of the long-range multipole expansion:\cite{Takagahara2000}
\begin{equation}\label{exchangedipole}
\begin{aligned}
S_{D(I)} \approx &\frac{1}{\pi\epsilon} \iint d^3\vec{r}\, d^3\vec{r}\,' \frac{\vec{\mu}^{\dagger}_{\uparrow \Downarrow}(\mathds{1}-\hat{n}\hat{n}^{\dagger})\vec{\mu}_{\downarrow \Uparrow}}{|\vec{r}-\vec{r}\,'|^3} \\
& \qquad \times \psi^{e*}_B(\vec{r}) \psi^e_B(\vec{r}\,') \psi^{h*}_{B(T)}(\vec{r}\,') \psi^h_{B(T)}(\vec{r}) ,
\end{aligned}
\end{equation}
where $\epsilon$ is the average permittivity of InAs and GaAs, $\mathds{1}$ is the $3\times 3$ identity matrix, $\hat{n}$ is a unit vector in the direction of $(\vec{r}-\vec{r}\,')$, and $\vec{\mu}_{\sigma \chi}=e\bra{u^e_{\sigma}}\hat{\vec{r}}\ket{u^h_{\chi}}$ is the interband transition dipole moment. Since $S_D$ varies widely in self-assembled QDs in the range $0-100$ $\mu$eV depending on detailed growth conditions, we leave it as a variable parameter in the model and numerically evaluate Eq.~(\ref{exchangedipole}) to obtain $S_I/S_D$ (see the appendix for details).

Incoherent transitions, described by operators $\hat{\sigma}_i$ and occuring at rates $\Gamma_i$, include optical recombination, phonon-assisted hole tunneling, spin-flip, and pure dephasing (see Table~\ref{incoherent}). These transitions are incoherent in the sense that they occur at random times, because we trace over the photonic and phononic environment degrees of freedom to obtain the reduced density matrix of the CQD system. We assume a direct exciton recombination rate $\Gamma_X=1~\mathrm{ns}^{-1}$ similar to experimentally observed rates in InAs-GaAs QDs. Indirect recombination rates $\Gamma_{iX}$ are found to be slower than direct recombination by a factor of $(M_{BT}/M_{BB})^2 \sim 100-1000$ due to a smaller electron-hole wavefunction overlap $M_{ij}=\left \langle \psi^e_i | \psi^h_j \right \rangle$ for the interdot state.\cite{BoyerdelaGiroday2011}

Holes undergo incoherent spin-conserving tunneling transitions via emission or absorption of a phonon, causing transitions between direct and indirect exciton states at rates given by Fermi's golden rule as
\begin{align}\label{Kabs}
\Gamma_{abs} &= \frac{2\pi}{\hbar} J(\Delta) n_B(\Delta,T) \\
\Gamma_{em} &= \frac{2\pi}{\hbar} J(\Delta) (n_B(\Delta,T) + 1),\label{Kem}
\end{align}
where $J(E)=\sum_{\vec{q}}|\bra{X}\hat{H}_{el-ph}\ket{iX}|^2 \delta(E_{\vec{q}}-E)$ is the phonon spectral density of the interdot transition, $n_B(E,T)=1/(e^{E/{k_B T}}-1)$ is the Bose distribution, giving the population of phonon modes at energy $E$ and temperature $T$, $\vec{q}$ denotes the phonon wave vector, $\hat{H}_{el-ph}$ is the electron-phonon interaction Hamiltonian, and we use a linear dispersion relation $E_{\vec{q}}=\hbar c_{LA} |\vec{q}|$ for acoustic phonons. We assume here that $\Delta \equiv E_X-E_{iX}>0$, so that tunneling to the direct exciton is partially suppressed, requiring absorption of a phonon. Since the detuning in the region of interest is much less than the LO phonon energy in GaAs, we consider only acoustic phonon-mediated transitions and neglect coupling to LO phonons. Including only interaction via the acoustic deformation potential, which is usually found to dominate for electronic transitions in quantum dots, leads to the expression\cite{Gawarecki2010}
\begin{equation}
J(\Delta)=\frac{\Delta a^2_V}{16\pi^3\rho c^2_{LA}} \int d^3\vec{q}\, \left| \int d^3\vec{r}\, \psi^*_{hB}e^{i\vec{q}\cdot\vec{r}} \psi_{hT} \right|^2 \delta(E_{\vec{q}}-\Delta) ,
\label{phononcoupling}
\end{equation}
where $a_V$ is the valence-band deformation potential, $\rho$ is the mass density of the crystal, $c_{LA}$ is the speed of the LA phonon mode in GaAs, and $\psi_{h\alpha}$ is the single-particle wavefunction for a hole localized in dot $\alpha=B,T$. We note that multiphonon and LO-phonon tunneling transitions are not included, which may become important at higher temperatures and exciton detunings than considered here.

Spin-flip transitions are also known to occur between exciton spin states, via phonon and spin-orbit coupling.\cite{Roszak2009,Tsitsishvili2010,Liao2011,Wei2014} In particular, spin relaxation occurs in direct (indirect) excitons primarily via phonon-assisted transitions between bright and dark spin states at a rate $\Gamma^{X(iX)}_{sf}$. Since our two-band effective mass model does not account for spin-orbit coupling, we include spin relaxation phenomenologically. We include only transitions between bright and dark spin states, since bright-bright and dark-dark transitions are at least an order of magnitude slower due to their smaller transition energies.\cite{Roszak2009} We use the temperature-dependence of the spin-flip rate measured by Fras et al.\cite{Fras2012}, scaled to fit the behavior measured by Hudson et al.\cite{Hudson2007}: $\Gamma^X_{sf}(T)=(0.27~\mathrm{ns}^{-1})+(0.29~\mu\mathrm{s}^{-1}\,\mathrm{K}^{-2})T^2$, corresponding to a thermally-activated two-phonon process. 
In the low-temperature regime where single-phonon transitions dominate and for small bright-dark splittings $\Delta^{D(I)}_{bd}$, $\Gamma^{X(iX)}_{sf}$ is given by Fermi's golden rule as in Eq's~(\ref{Kabs}) and (\ref{Kem}), with $J(\Delta^{D(I)}_{bd})\propto(\Delta^{D(I)}_{bd})^3$ and $n(\Delta^{D(I)}_{bd},T)\approx k_B T/ \Delta^{D(I)}_{bd}$.\cite{Eastham2013} Since the bright-dark splitting scales with electron-hole wavefunction overlap, we predict the relationship $\Gamma^{iX}_{sf}=(M_{BT}/M_{BB})^2 \,\Gamma^X_{sf}$ between direct and indirect spin-flip rates.

\begin{figure}
\centering
\includegraphics[width=0.48\textwidth]{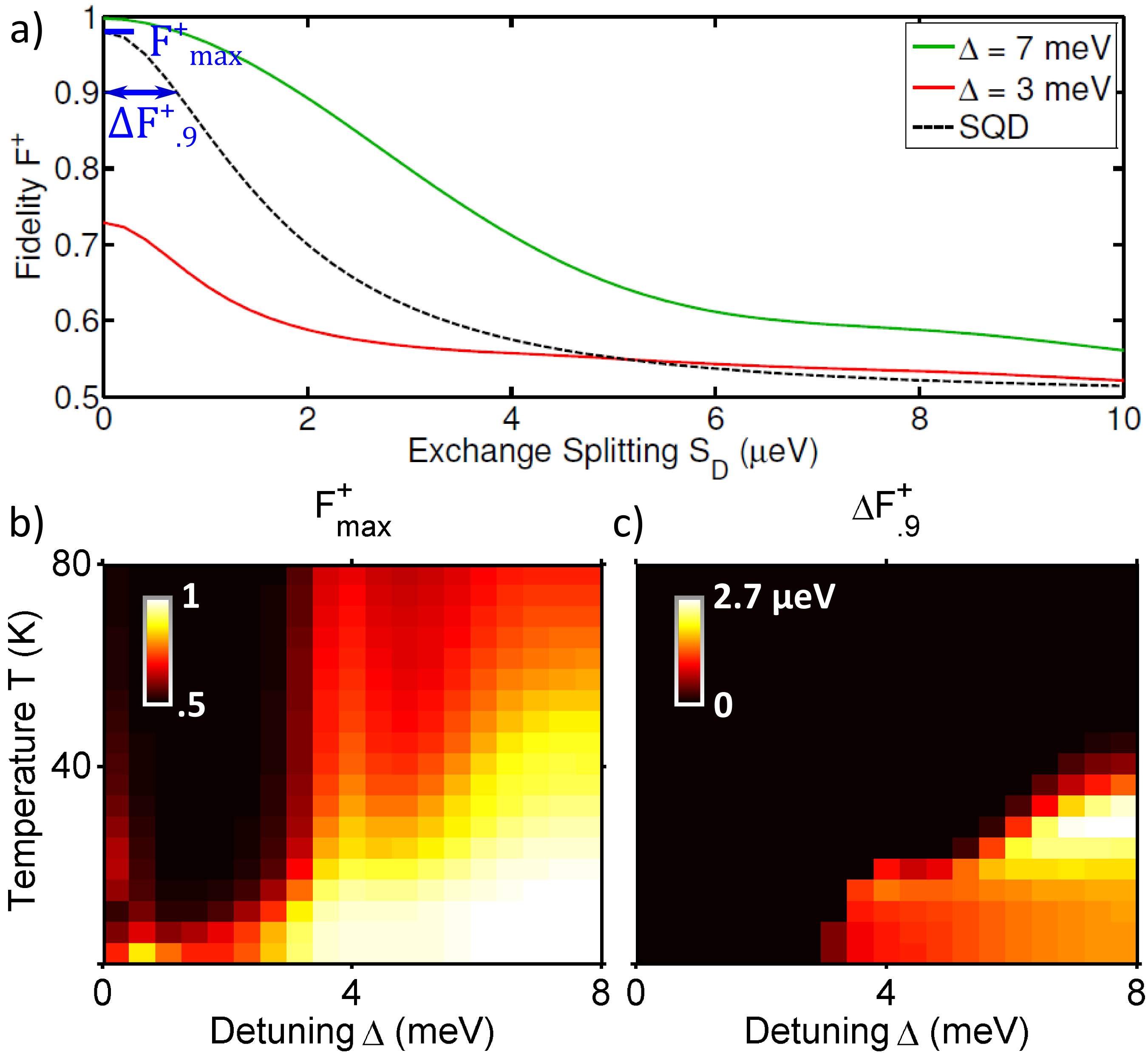}
\caption{(a) Dependence of entanglement fidelity on direct exciton bright state splitting at $T=20~$K, $t_h=0.2~\mu$eV, and $\Delta=7~$meV (green, top curve) and $\Delta=3~$meV (red, bottom curve), compared to the analogous SQD case. (b) Maximum fidelity and (c) width of $F^+(S_D)$ curve as a function of exciton detuning and temperature.}
\label{fidelityresults}
\end{figure}

Phonon-induced pure dephasing can have a substantial effect on interband coherences, most notably an increase of transition linewidth at elevated temperatures.\cite{Borri2005} Since phonons carry zero spin however, they couple equally to each electronic spin state. While, in principle, a slight spin-dependence of the phonon coupling could lead to a pure cross-dephasing $\gamma^{iX}_{cd}$ of the indirect exciton coherence,\cite{Hudson2007} we neglect this possibility for the remainder of this work on the grounds that Moody et al.\cite{Moody2013} have measured the cross-dephasing rate to be negligible in the case of a single QD. As a result, the relative phase between exciton spin states---and the resulting photon entanglement---is largely unaffected by phonon-induced pure dephasing processes.

\section{Entanglement}\label{entanglement}

The results of the numerical simulation are summarized in Fig.~\ref{fidelityresults}. The fidelity approximately exhibits a Lorentzian dependence on the direct exciton exchange splitting $S_D$, similar to the behavior reported in the case of a single QD.\cite{Hudson2007,Carmele2010,Schumacher2012,Trotta2014,Zhang2015} To characterize this behavior, we record the maximum fidelity $F^+_{max}$ at $S_D = 0$ and the fidelity width $\Delta F^+_{.9}$, defined as the largest value of $|S_D|$ with $F^+>.9$. By plotting these fidelity characteristics as a function of the experimentally tunable exciton detuning $\Delta$ and temperature $T$, we obtain a map of possible behaviors observable in a single CQD pair. In general, we observe that the quality of entanglement depends primarily on the total dephasing rate
\begin{equation}
\gamma^{iX}_d = \Gamma_{iX}+\Gamma_{abs}+\Gamma^{iX}_{sf}
\end{equation}
describing decay of the indirect exciton bright state coherence $\bra{iX,+1}\hat{\rho}\ket{iX,-1}$, as well as the effective decay rate $\Gamma_{eff}$ of the indirect exciton population $\bra{iX}\hat{\rho}\ket{iX}$, determined by an exponential fit of the bright state density matrix dynamics. For low temperatures ($T<40~$K) and large detunings ($\Delta>3~$meV), entanglement fidelity can reach values above $0.99$ for small exchange splittings, with a width $\Delta F^+_{.9} \approx 1.5-2.7~\mu$eV. Note that the CQD system exhibits this high-fidelity behavior with $S_D$ values up to a factor of 5 greater than in the single QD case ($\Delta F^+_{.9,SQD} \approx 0.4-0.6~\mu$eV). In this region, the decoherence is radiatively-limited ($\gamma^{iX}_d<2\,\Gamma_{iX}$), allowing coherent evolution between photon emission events. With increasing lattice temperature, the overall fidelity decreases due to phonon-assisted tunneling and spin-flip. At detuning values below $3~$meV, the large dephasing rate ($\gamma^{iX}_d>100\,\Gamma_{iX}$) prevents photon entanglement even in the absence of anisotropic exchange splitting, giving a maximum fidelity near $0.5$.

\begin{figure}
\centering
\includegraphics[width=0.48\textwidth]{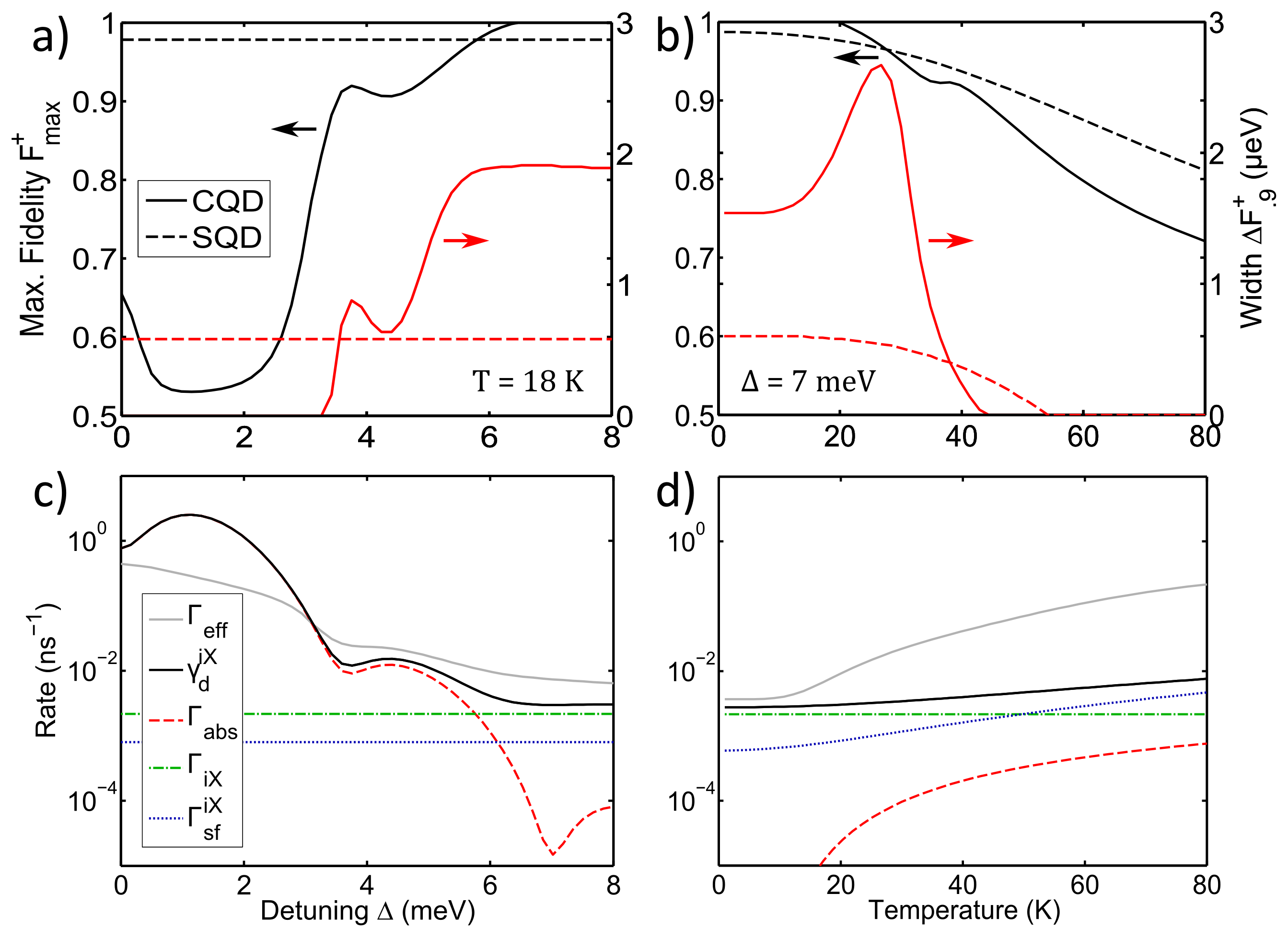}
\caption{(a) (b) Dependence of fidelity parameters $F^+_{max}$ (black, left y-axis) and $\Delta F^+$ (red, right y-axis) on (a) detuning and (b) temperature, with $t_h=0.2~\mu$eV. (c) (d) Total dephasing rate $\gamma^{iX}_d$ of the indirect exciton bright state coherence, including contributions from incoherent transitions, as a function of detuning and temperature, respectively. Also included is the effective $\ket{iX}$ population decay rate $\Gamma_{eff}$.}
\label{transitionrates}
\end{figure}

Fig's~\ref{transitionrates}(a) and (b) isolate the effect of exciton detuning and temperature, respectively, on the different fidelity characteristics. Fig's~\ref{transitionrates}(c) and (d) show the dephasing rate $\gamma^{iX}_d$ over the same ranges, along with each of the contributing transition rates and the effective indirect exciton population decay rate $\Gamma_{eff}$. As in Fig.~\ref{fidelityresults}, regions of high fidelity correlate with a radiatively-limited dephasing rate. The significant drop in fidelity at low exciton detunings is therefore due to a phonon-assisted hole tunneling rate which surpasses radiative emission by up to 3 orders of magnitude. Note that the oscillations in the tunneling rates are due to the phase $e^{i\vec{q}\cdot\vec{r}}$ appearing in the phonon coupling (Eq. (\ref{phononcoupling})), and depend on the distance $d$ between QDs.\cite{Rolon2012} The peak in fidelity width as a function of temperature occurs when $\Gamma_{eff}$ increases, while $\gamma^{iX}_d$ remains low. The increased decay rate of the indirect exciton causes the time-integration in Eq. (\ref{tomography}) to sample shorter time delays, maintaining high fidelity over a larger range of exchange splittings. While hole tunneling can be adequately suppressed by simply maintaining a large enough detuning between exciton states or increasing the tunnel barrier, phonon-assisted spin-flip surpasses radiative emission and reduces fidelity at temperatures above $\sim 50~$K.

\begin{figure}
\centering
\includegraphics[width=0.48\textwidth]{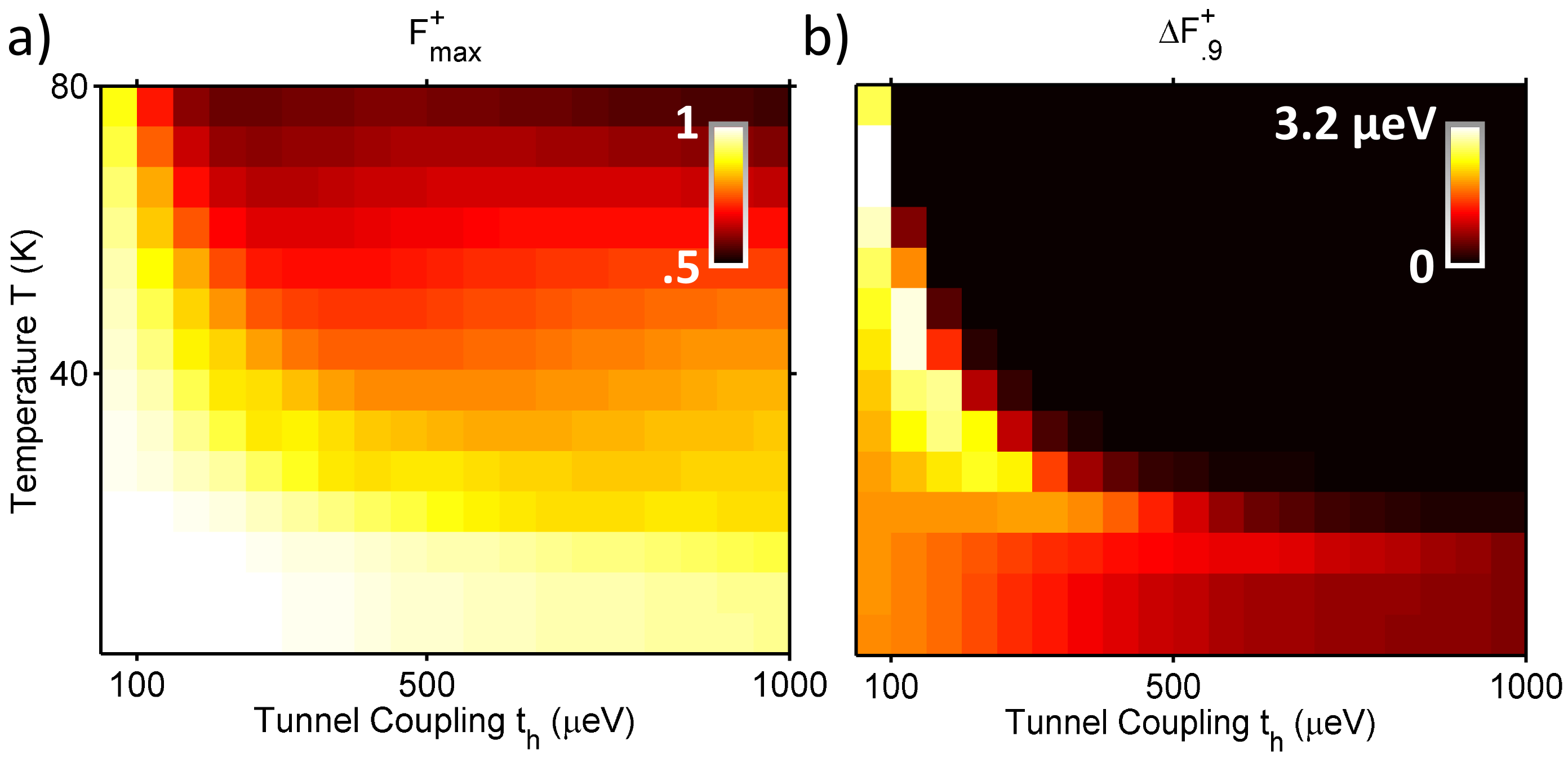}
\caption{(a) Maximum fidelity and (b) width of $F^+(S_D)$ curve as a function of hole tunnel coupling and temperature, with $\Delta = 7~$meV.}
\label{fidelity_tunnelcoupling}
\end{figure}

Fig.~\ref{fidelity_tunnelcoupling} shows the fidelity characteristics as a function of hole tunnel coupling and temperature, at a fixed exciton detuning. The effective indirect exciton decay rate $\Gamma_{eff}$ depends strongly on both temperature and tunnel coupling, and is demonstrated to have a substantial effect on the entanglement fidelity. $F^+_{max}$ remains higher than $0.9$ where $\Gamma_{eff}<30\,\Gamma_{iX}$, which occurs at low temperature and tunnel coupling values. The fidelity width $\Delta F^+_{.9}$ peaks at intermediate values of $\Gamma_{eff}$ ($\Gamma_{eff}\approx 5-10\,\Gamma_{iX}$) and increases with decreasing tunnel coupling, reaching values above $3.0~\mu$eV at small tunnel couplings ($t_h\leq .1~$meV) and temperatures up to $70~$K. As tunnel coupling increases, the indirect exciton gains more of a direct exciton character, decreasing the fidelity width and requiring lower temperatures to suppress recombination.

Our results indicate that by maintaining a large enough exciton detuning to suppress phonon-assisted tunneling, CQD-based entangled photon sources can produce entangled photon pairs with higher fidelity than single QD-based sources, over a wider range of direct exciton exchange splittings. While this helps reduce the strict symmetry requirement for entanglement generation, the photons are separated by a relatively long time delay of $100-1000~$ns due to the low indirect exciton recombination rate. In the intermediate state, however, the indirect exciton spin is entangled with the polarization of the first emitted photon (Eq.~\ref{intermediatestate}). As a result, this scheme could potentially be used to entangle spins in remote CQDs by tuning each of the first emitted photons into resonance via electric or strain fields and performing a joint polarization correlation measurement.\cite{Simon2003,Simon2007,Gao2012}

\section{Conclusion}

We have used a simple theoretical model to simulate the radiative cascade of the molecular biexciton state in a vertically stacked tunnel-coupled quantum dot pair. The entanglement fidelity of the resulting two-photon polarization state is determined, accounting for phonon-assisted tunneling and spin-flip processes. From numerical simulations, we find an approximately Lorentzian dependence of fidelity on anisotropic electron-hole exchange splitting and mapped the behavior over a range of electric field and temperature values. Our results show that near-unity maximum fidelity can be achieved over a range of exchange splittings $|S_D|<2.7~\mu$eV at large exciton detunings and low temperatures, where dephasing due to phonon-assisted hole tunneling and spin-flip processes is suppressed. This suggests that coupled quantum dots can generate photon pairs with a high degree of entanglement and over a wider range of exchange splittings compared to single dots, provided the tunnel coupling is low enough to maintain charge separation in the indirect exciton state. In addition, the spin-photon entanglement generated by the first recombination could be used to entangle spins in remote CQDs.

\section*{Acknowledgements}
We are grateful to S. Economou for helpful discussions. This work was supported by the University of California, Merced. One of us (M.S.) acknowledges funding from the Hellman Family Faculty Foundation.

\appendix

\section*{Appendix: Computational Methods}
Electron and hole single-particle wavefunctions are calculated using the two-band effective mass approximation. We model the confinement potential of each QD as a finite well in the growth (z) direction and parabolic confinement in the lateral (x-y) directions, expressed for particle $\alpha=e,h$ as $V_{\alpha}(\vec{r})=\frac{1}{2}m_{\alpha,\mathrm{InAs}}\omega_{\alpha}^2(x^2+y^2) + \mathcal{E}_{\alpha}\theta(|z|-\frac{h}{2})$, where $m_{\alpha,\mathrm{InAs}}$ is the effective mass of particle $\alpha$ in InAs, $\omega_{\alpha}$ is the angular frequency of the lateral harmonic oscillator, and $\mathcal{E}_{\alpha}$ is the conduction/valence band offset of the heterostructure. Envelope wavefunctions of single-particle localized ground states are then defined on a Cartesian grid as
\begin{equation}
\begin{aligned}
\psi_{\alpha}(\vec{r}) = & A e^{-\frac{m_{\alpha,\mathrm{InAs}}\omega_{\alpha}}{2\hbar}(x^2+y^2)} \times \\
& \times
\begin{cases}
\cos(kh/2) e^{-\kappa(z - h/2)} & \text{if } z > h/2 \\
\cos(kz) & \text{if } |z| < h/2 \\
\cos(kh/2) e^{\kappa(z + h/2)} & \text{if } z < -h/2 ,
\end{cases}
\end{aligned}
\label{wavefunction}
\end{equation}
where the wavenumber $k$ is determined by the transcendental equation $\tan(kh/2)=\sqrt{\frac{m_{\alpha,\mathrm{InAs}}}{m_{\alpha,\mathrm{GaAs}}}\left(\frac{k_0^2}{k^2}-1\right)}$ due to the boundary conditions, $\kappa=k\sqrt{\frac{k_0^2}{k^2}-\frac{m_{\alpha,\mathrm{GaAs}}}{m_{\alpha,\mathrm{InAs}}}}$, $k_0=\sqrt{2m_{\alpha,\mathrm{InAs}}\mathcal{E}_{\alpha}}/\hbar$, and $A$ is a normalization constant defined such that $\int |\psi_{\alpha,B/T}(\vec{r})|^2 \, d^3\vec{r} = 1$, and $h$ is the QD height in the growth direction. In a coordinate system with the origin set at the center of the two QDs, wavefunctions for particles localized in each dot are found from Eq.~\eqref{wavefunction} as $\psi_{\alpha,B/T}(x,y,z)=\psi_{\alpha}(x,y,z \pm d/2)$, with the substitution $h \mapsto h_{B/T}$. The energies of these single-particle ground states are then given by
\begin{equation}
E_{\alpha,B/T} = \frac{\hbar^2 k^2}{2 m_{\alpha,\mathrm{InAs}}} + \hbar\omega_{\alpha}.
\end{equation}

\begin{table}
\caption{List of material and CQD parameters used in numerical simulations.}
\setlength{\extrarowheight}{3pt}
\begin{tabular}{c c c}
\hline
\hline
Material Parameters & & \\
\hline
Permittivity & $\epsilon$ & $13.5~\epsilon_0$ \\
InAs electron mass & $m_{e,\mathrm{InAs}}$ & $0.059~m_0$ \\
GaAs electron mass & $m_{e,\mathrm{GaAs}}$ & $0.042~m_0$ \\
InAs hole mass & $m_{h,\mathrm{InAs}}$ & $0.37~m_0$ \\
GaAs hole mass & $m_{h,\mathrm{GaAs}}$ & $0.34~m_0$ \\
Strained InAs band gap & $E_g$ & $866~$meV \\
Conduction band offset & $\mathcal{E}_e$ & $461~$meV \\
Valence band offset & $\mathcal{E}_h$ & $192~$meV \\
Hole deformation potential & $a_V$ & $700~$meV \\
Mass density & $\rho$ & $5300~\mathrm{kg}/\mathrm{m}^3$ \\
Speed of sound & $c_{LA}$ & $5150~$m/s \\
\hline
CQD Parameters & & \\
\hline
Bottom QD height & $h_B$ & $3.0~$nm \\
Top QD height & $h_T$ & $2.5~$nm \\
QD separation & $d$ & $7.0~$nm \\
Electron level spacing & $\hbar \omega_e$ & $66~$meV \\
Hole level spacing & $\hbar \omega_h$ & $15~$meV \\
Hole tunnel coupling & $t_h$ & $200~\mu$eV \\
Bright-dark splitting & $\Delta_{bd}$ & $300~\mu$eV \\
Direct exciton recombination rate & $\Gamma_X$ & $1.0~\mathrm{ns}^{-1}$ \\
\hline
\hline
\end{tabular}
\label{parameters}
\end{table}

The Coulomb interaction energy between particles $\alpha$ and $\beta$ in dots $i$ and $j$, respectively, is given by
\begin{equation}
V^{\alpha \beta}_{ij}=\frac{e^2}{4\pi\epsilon} \iint d^3\vec{r}\, d^3\vec{r}\,'\, \frac{|\psi_{\alpha i}(\vec{r})|^2 |\psi_{\beta j}(\vec{r}\,')|^2}{|\vec{r}-\vec{r}\,'|},
\end{equation}
where $e$ is the charge of an electron, $\epsilon$ is the average permittivity of InAs and GaAs, and the 6-dimensional integral is directly evalutated by numerical integration over a Cartesian grid. The Coulomb interaction terms modify the energy of multiparticle exciton states, as does the local electric field $F$ due to the large permanent dipole moment $p=ed$ of indirect exciton states. The energies of the various charge states considered are then
\begin{equation}
\begin{aligned}
E_X =& E_g + E_{eB} + E_{hB} - V^{eh}_{BB} \\
E_{iX} =& E_g + E_{eB} + E_{hT} - V^{eh}_{BT} - edF \\
E_{iXX} =& 2E_g + 2E_{eB} + E_{hB} + E_{hT} \\
&+ V^{ee}_{BB} + V^{hh}_{TT} - 2V^{eh}_{BB} - 2V^{eh}_{BT} - edF ,
\end{aligned}
\end{equation}
where $E_g$ is the band gap of the strained InAs comprising each QD.

Eq.~\eqref{exchangedipole} is expanded using the interband dipole moments $\vec{\mu}_{\uparrow\Downarrow(\downarrow\Uparrow)} = \mu (\pm 1, -i, 0)/\sqrt{2}$ to obtain the expression
\begin{equation}
\begin{aligned}
S_{D(I)} = \frac{\mu^2}{2\pi\epsilon} \iint & d^3\vec{r}\, d^3\vec{r}\,'\, \frac{2\Delta z^2 - \Delta x^2 - \Delta y^2 + 6i\Delta x \Delta y}{\Delta r^5} \\
& \times \psi^*_{eB}(\vec{r}) \psi_{eB}(\vec{r}\,') \psi^*_{hB(T)}(\vec{r}\,') \psi_{hB(T)}(\vec{r}),
\label{exchangesplitting}
\end{aligned}
\end{equation}
where $\Delta\vec{r}=\vec{r}-\vec{r}\,'$. Since the exchange splitting varies widely between QDs and small values are required for entanglement generation, we choose values of $S_D$ and numerically integrate Eq.~\eqref{exchangesplitting} to obtain the ratio $S_I/S_D$.

We use the value of $\Gamma_X$ listed in Table~\ref{parameters} and the electron-hole wavefunction overlap
\begin{equation}
M_{ij}=\int d^3\vec{r}\, \psi_{ei}^*(\vec{r}) \psi_{hj}(\vec{r})
\end{equation}
obtained by numerical integration to calculate $\Gamma_{iX}=\Gamma_X (M_{BT}/M_{BB})^2$. The phonon-assisted tunneling rates $\Gamma_{abs}$ and $\Gamma_{em}$ are calculated using Eq's \ref{Kabs}, \ref{Kem} and \ref{phononcoupling}, where the phonon wave vector $\vec{q}$ is expressed in spherical coordinates and its magnitude is constrained by the delta function to be $q=\Delta/\hbar c_{LA}$. To calculate the phonon-assisted spin-flip rate of the direct exciton, we use the experimentally-determined temperature dependence $\Gamma^X_{sf}=(.27~\mathrm{ns}^{-1}) + (.29~\mu\mathrm{s}^{-1}\mathrm{K}^{-2})T^2$, and from it determine the indirect exciton spin-flip rate $\Gamma^{iX}_{sf}=\Gamma^X_{sf} (M_{BT}/M_{BB})^2$. For the pure dephasing rate, we use the empirical temperature dependence $\hbar\gamma^X_p=(71~\mu\mathrm{eV})n_B(6~\mathrm{meV},T) + (4.5~\mathrm{meV})n_B(28~\mathrm{meV},T)$ corresponding to phonon-assisted pure dephasing processes involving excited hole states, and assume the same pure dephasing rate for indirect excitons: $\gamma^{iX}_p=\gamma^X_p$.

By projecting Eq.~\ref{masterequation} onto each combination of basis states, it becomes $\dot{\rho}_{ij}=\sum_{kl} M_{ij,kl}\rho_{kl}$ and we determine the elements of the time-dependence tensor $M_{ij,kl}$. We then vectorize $\hat{\rho}$ by mapping the matrix elements $\rho_{ij}$ to a one-dimensional vector $\vec{\rho}$. The time-dependence tensor $M_{ij,kl}$ is then transformed into a matrix $M_{ij}$, and the solution is expressed by a matrix exponential as $\vec{\rho}(\tau)=e^{\mathcal{L}\tau}\vec{\rho}(0)=e^{Mt}\vec{\rho}(0)$. With this matrix exponential describing time evolution of the CQD density matrix, we use Eq's~\ref{correlations} and \ref{tomography} to calculate the various elements of the two-photon polarization density matrix $\hat{\rho}_{2P}$ in the linear polarization basis $\{\ket{HH},\ket{HV},\ket{VH},\ket{VV}\}$, averaging over delay times up to $T_d=200~$ns. With these density matrix elements, the fidelity to the entangled Bell state is finally calculated using Eq.~\ref{fidelity}.

\bibliographystyle{apsrev}

\end{document}